# A critical comparative analysis of five world university rankings

Henk F. Moed[1]

Sapienza University of Rome, Italy

## Summary

To provide users insight into the value and limits of world university rankings, a comparative analysis is conducted of 5 ranking systems: **ARWU**, **Leiden**, **THE**, **QS** and **U-Multirank**. It links these systems with one another at the level of individual institutions, and analyses the overlap in institutional coverage, geographical coverage, how indicators are calculated from raw data, the skewness of indicator distributions, and statistical correlations between indicators. Four secondary analyses are presented investigating national academic systems and selected pairs of indicators. It is argued that current systems are still one-dimensional in the sense that they provide finalized, seemingly unrelated indicator values rather than offer a dataset and tools to observe patterns in multi-faceted data. By systematically comparing different systems, more insight is provided into how their institutional coverage, rating methods, the selection of indicators and their normalizations influence the ranking positions of given institutions.

## 1. Introduction

In most OECD countries, there is an increasing emphasis on the effectiveness and efficiency of government-supported research. Governments need systematic evaluations for optimizing their research allocations, re-orienting their research support, rationalizing research organizations, restructuring research in particular fields, or augmenting research productivity. In view of this, they have stimulated or imposed evaluation activities of their academic institutions. Universities have become more diverse in structure and are more oriented towards economic and industrial needs.

In March 2000, the European Council agreed a new strategic goal to make Europe "the most competitive and dynamic knowledge-based economy in the world, capable of sustainable economic growth with more and better jobs and greater social cohesion". Because of the importance of research and development to "generating economic growth, employment and social cohesion", the Lisbon Strategy says that European universities "must be able to compete with the best in the world through the completion of the European Higher Education Area" (EU Council, 2000). In its resolution 'Modernizing Universities for Europe's Competitiveness in a Global Knowledge Economy', the European Council expressed the view that the "challenges posed by globalization require that the European Higher Education Area and the European Research Area be fully open to the world and that Europe's universities aim to become worldwide competitive players" (EU Council, 2007, p. 3).

An Expert Group on the assessment of university-based research noted in 2009 that university rankings have become an increasing influence on the higher education landscape since US News and World Report began providing consumer-type information about US universities in 1983. They "enjoy a high level of acceptance among stakeholders and the wider public because of their simplicity and consumer type information" (AUBR Expert Group, 2009, p 9).

---

[1] Visiting professor. Email: henk.moed@uniroma1.it; hf.moed@gmail.com

University ranking systems have been intensely debated, for instance by Van Raan (2005), Calero-Medina et al. (2008), Salmi (2009), Hazelkorn (2011), Rauhvargers (2011; n.d.) and Shin, Toutkoushian and Teichler (eds.) (2011). A report from the European University Association concluded that despite their shortcomings, evident biases and flaws, rankings are here to stay. "For this reason it is important that universities are aware of the degree to which they are transparent, from a user's perspective, of the relationship between what it is stated is being measured and what is in fact being measured, how the scores are calculated and what they mean" (Rauhvargers, 2011, p. 7).

A base notion underlying the current article is that a critical, *comparative* analysis of a *series* of university ranking systems can provide useful knowledge that helps a wide range of interested users to better understand the information provided in these systems, and to interpret and use it in an informed, responsible manner. The current article aims to contribute to such an analysis by presenting a study of the following five ranking systems: *ARWU* World University Rankings 2015, CWTS *Leiden* Ranking 2016, *QS* World University Rankings 2015-2016, *THE* World University Rankings 2015-2016, and *U-Multirank* 2016 Edition. An overview of the indicators included in the various systems is given in Table A1 in the Appendix.

*ARWU*, the Academic Ranking of World Universities, also indicated as 'Shanghai Ranking' is the oldest ranking system. Initially created by the Center for World-Class Universities (CWCU) at Shanghai Jiao Tong University, since 2009 it has been published and copyrighted by ShanghaiRanking Consultancy. It combines bibliometric data from Thomson Reuters with data on prizes and awards of current and former academic staff or students. The *ARWU* 2015 Ranking of World Universities, available online and analyzed in the current article, covers 500 institutions. The *Leiden* Ranking is not a ranking in the strict sense but rather a bibliometric information system, containing for about 850 universities bibliometric data extracted from Web of Science related to publication output, citation impact and scientific collaboration. This article uses the 2016 version of the database.

*U-Multirank* is prepared with seed funding from the European Union by a Consortium lead in 2016 by the Center for Higher Education Policy Studies (CHEPS), The Netherlands; Centre for Higher Education (CHE) in Germany; and the Centre for Science and Technology Studies (CWTS), Leiden University, The Netherlands. This article is based on the 2016 version. A key feature of the *U-Multirank* system is the inclusion of teaching and learning-related indicators. While some of these relate to a university as a whole, the core part is concerned with 13 specific scientific-scholarly disciplines, and based on a survey among students.

Between 2004 and 2009, **Times Higher Education (*THE*)** and **Quacquarelli Symonds (*QS*)** jointly published the *THES-QS* World University Rankings. After they had ended their collaboration, the methodology for these rankings continued to be used by *QS* as the owner of its intellectual property. Since 2010 these rankings are known as the *QS* World University Rankings. At the same time, *THE* started publishing another ranking, applying a methodology developed in partnership with Thomson Reuters in 2010, known as the Times Higher Education or *THE* World University Rankings and related rankings. At present, both organizations have a collaboration with Elsevier, and use bibliometric data from Scopus.

A series of interesting studies analysed statistical properties and validity *within* particular university ranking systems (e.g., Soh, 2013; Paruolo, Saisana and Saltelli, 2013; Soh, 2015a; Soh, 2015b;), mostly focusing on the so called *Overall* indicator which is calculated as a weighted sum of the various indicators. For instance, a factor analysis per ranking system conducted by Soh (2015a) found that the factors identified in **ARWU**, **THE** or **QS** systems are negatively correlated or not correlated at all, providing evidence that the indicators covered by each system are not "mutually supporting and additive". Rather than dealing with the internal consistency and validity *within* a particular system, the current paper makes comparisons *among* systems.

All five systems listed above claim to provide valid and useful information for determining academic excellence, and have their own set of indicators for measuring excellence. Three systems, **ARWU**, **THE** and **QS**, present an overall indicator, by calculating a weighted sum of scores of a set of key indicators. The **Leiden** Ranking and **U-Multirank** do not have this type of composite measure. The current paper examines the consistency among the systems. As all systems claim to measure essentially academic excellence, one would expect to find a substantial degree of consistency among them. The overarching issue addressed in the current paper is the assessment of this consistency-between-systems. To the extent that a lack of consistency is found, – and the next chapters will show that it exists –, what are the main causes of the observed discrepancies? What are the systems' profiles? How can one explain to potential users the ways in which the systems differ one from another? What are the implications of the observed differences for the interpretation and use of a particular system as a 'stand-alone' source of information?

The article consists of two parts. In the *first* part, a series of statistical properties of the 5 ranking systems are analyzed. The following research questions are addressed.

- *Overlap in institutional coverage (Section 2).* How many institutions do the rankings have pairwise in common? And what is the overlap between the top 100 lists in the various rankings? If this overlap is small, one would have to conclude that the systems have different ways to define academic excellence, and that it is inappropriate to speak of "*the*" 100 global top institutions.
- *Differences in geographical coverage (Section 3).* How are the institutions distributed among countries and world regions in which they are located? Are there differences in this distribution between ranking systems? All five systems claim to adopt a global viewpoint; **ARWU**, **THE** and **QS** explicitly speak of *world* universities. But do they analyse the world in the same manner? Are differences between global geographical regions mainly due to differences in excellence in those regions, or do regional indicator normalizations play a significant role as well?
- *Indicator distributions and their skewness (Section 4).* Firstly, to which extent do the systems present for each institution they cover scores for *all* indicators? When assessing the information content of a system, it is important to have an estimate of the frequency of occurrence of missing values. Secondly, which methods do the systems apply to calculate scores from the raw data? Such methods determine how differences in indicator scores should be interpreted in terms of differences in underlying data. For instance, **ARWU**, **THE** and **QS** express an indicator score as a number ranging from 0 to 100, while U-Multirank uses five so called performance classes (A to E). How precisely are these scores defined, and, especially, which differences exist between systems? Finally, how does the skewness of indicator distributions vary between indicators and between ranking systems? To what extent is skewness as measured by the various systems a base characteristic of the global academic system, or is it determined by the way in which the systems calculate their indicators?

- *Statistical correlations between indicators (Section 5)*. The least one would expect to find when comparing ranking systems is that (semi-) identical indicators from different systems, such as the number of academic staff per student, show a very strong, positive correlation. Is this actually the case? Next, how do indicators from different systems measuring the same broad aspect (e.g., citation impact or academic reputation) correlate? If the correlation is low, what are the explanations? To what extent are indicators complementary?

In the *second* part of the paper (Section 6) four analyses show how a more detailed analysis of indicators included in a system, and, especially, how the *combination* of indicators from *different* systems can generate useful, new insights and a more comprehensive view on what indicators measure. The following analyses are presented.

- *Characteristics of national academic systems.* What is the degree of correlation between citation- and reputation-based indicators in major countries? This analysis is based on indicators from the *THE* ranking. It aims to illustrate how simple data representations, showing for instance in scatterplots how pairs of key indicators for a given set of institutions are statistically related, can provide users insight into the structure of underlying data, raise critical questions, and help interpreting the indicators.
- *QS* versus *Leiden* *citation-based indicators.* What are the main differences between these two indicators? How strongly do the correlate? Are they interchangeable? The main purpose of this analysis is to show how indicator normalization can influence the rank position of given universities, and also to underline the need to systematically investigate the data quality of 'input-like' data such as number of students or academic staff obtained via institutional self-reporting or from national statistical offices.
- *THE* *Research Performance versus* *QS* *Academic Reputation.* What are the main differences between the *THE* and *QS* reputation-based indicators? How strongly do they correlate? Which institutions show the largest discrepancies between *THE* and *QS* score? This analysis provides a second illustration of how indicator normalization influences university rankings.
- *ARWU* *Highly Cited Researchers vs.* *Leiden* *Top Publications indicator.* Gingras (2014) found severe biases in the Thomson Reuters List of Highly Cited Researchers, especially with respect to Saudi Arabian institutions. Do these biases affect the *ARWU* indicator that uses this list as data source? This fourth study shows how a systematic comparison of indicators of the same broad aspect from different systems can help interpreting the indicators, and evaluating their data quality and validity.

Finally, Section 7 presents a discussion of the outcomes and makes concluding remarks.

## 2. Analysis of institutional overlap

In a first step, data on the names and country of location of all institutions, and their values and rank positions for all indicators in as far as available were extracted from the websites of the 5 systems, indicated in Table A1 in the Appendix. Next, names of institutions were standardized, by unifying major organizational and disciplinary terms (e.g., 'university', 'scientific') and city names (e.g., 'Roma' vs. 'Rome'), and an initial version of a thesaurus of institutions was created, based on their appearance in the first ranking system. Next, this thesaurus was stepwise expanded, by matching it against the institutional names from a next ranking system, manually inspecting the results, and updating it, adding either new variant names of institutions already included, or names of new institutions not yet covered. As a final check, names of institutions appearing in the top 100 of one system but not found in the other systems, were checked manually. In the end, 1,715 unique institutions were identified, and 3,248 variant names. 377 universities (22 per cent) appear in all 5 ranking systems, and 182 (11 per cent) in 4 systems.

A major problem concerning university systems in the USA was caused by the fact that it was not always clear which components or campuses were covered. For instance, University of Arkansas System has 6 main campuses. **ARWU** has two entries, 'U Arkansas at Fayetteville' and 'U Arkansas at Little Rock'. **Leiden** includes 'U Arkansas, Fayetteville' and 'U Arkansas for Medical Sciences, Little Rock'. **QS**, **THE**, and **U-Multirank** have one entry only, 'U Arkansas'. Similar problems occur for instance with 'Univ Colorado', 'Univ Massachusetts', 'Purdue Univ' and 'Univ Minnesota'. If it was unclear whether two institutions from different ranking systems covered the same components or campuses, they were considered as different, even if there is a substantial overlap between the two.

Table 1: Institutional overlap between the 5 ranking systems

|  | ARWU | LEIDEN | QS | THE | U-MULTIRANK |
|---|---|---|---|---|---|
| ARWU | 500 | 468 | 444 | 416 | 465 |
| LEIDEN |  | 840 | 585 | 589 | 748 |
| QS |  |  | 917 | 635 | 638 |
| THE |  |  |  | 800 | 627 |
| U-MULTIRANK |  |  |  |  | 1,293 |

Table 1 presents the institutional overlap between each pair of ranking systems. The numbers in the diagonal give the total number of institutions covered by a particular system. Table 2 gives key results for the overlap in the top 100 lists of all 5 systems. It shows that the total number of unique institutions in the top 100 lists of the five systems amounts to 194. Of these, 35 appear in all lists.

Table 2. Key results overlap analysis of top 100 lists in all 5 ranking systems

| Indicator | N |
|---|---|
| Total number of different institutions | 194 |
| Number of institutions appearing in the top 100 lists of all 5 systems | 35 |

Table 3 shows the institutional overlap between *the top 100 lists* of the various systems. For **ARWU**, **QS** and **THE** the 'overall', weighted ranking was used. **Leiden** and **U-Multirank** do not include such an overall ranking. For **Leiden**, two top 100 lists were created, one size-dependent, based on the number of publications (labelled as LEIDEN-PUB in Table 3), and a second size-independent (LEIDEN-CIT), based on the Mean Normalized Citation Score (MNCS), a size-normalized impact measure correcting for differences in citation frequencies between subject fields, the age of cited publications, and their publication type (see Leiden Indicators, n.d.). Since there is no obvious preferred ranking in **U-Multirank**, this system was not included in Table 3. The number of overlapping institutions per pair of systems ranges between 49 for the overlap between the two **Leiden** top lists, and 75 for that between **QS** and **THE**.

Table 3: Institutional overlap between the top 100 lists of 4 ranking systems

|  | LEIDEN-CIT | LEIDEN-PUB | QS | THE |
|---|---|---|---|---|

| ARWU | 60 | 67 | 60 | 66 |
|---|---|---|---|---|
| LEIDEN-CIT | | 49 | 51 | 56 |
| LEIDEN-PUB | | | 64 | 68 |
| QS | | | | 75 |

It should be noted that the overwhelming part of the top institutions in one ranking but missing in the *top 100* of another ranking were found at *lower* positions of this other ranking. In fact, the number of cases in which a top institution in a system is not linked to any university in another system ranges between 0 and 6, and most of these relate to institutions in university systems located in the USA.

Several cases were detected of institutions that could not be found in a system, while one would expect them to be included on the basis of their scores in other systems. For instance, *Rockefeller University,* occupying the 33th position in the overall ***ARWU*** ranking, and first in the ***Leiden*** ranking based on normalized citation rate, is missing in the ***THE*** ranking. *Freie Univ Berlin* and *Humboldt Univ Berlin* – both in the top 100 of the overall ***THE*** ranking and in the top 150 of the ***QS*** ranking – could not be found in the ***ARWU*** system, while *Technical Univ Berlin*, ranking 178[th] in the ***QS*** system, was not found in the ***THE*** system. In the ***THE*** *World Ranking* the Italian institutions *Scuola Normale Superiore di Pisa* and *Scuola Superiore Santa Anna* are in the range 101-200. In fact, the first has the largest score on the ***THE*** Research Performance indicator. But institutions with these two names do *not* appear in the ***QS*** *World University Ranking*; it is unclear whether the entity '*University of Pisa'*, appearing in the overall ***QS*** ranking on position 367, includes these two schools.

## 3. Geographical distributions

The preference of ranking system R for a particular country C is expressed as the ratio of the actual and the expected number of institutions from C appearing in R, where the expected number is based on the total number of institutions across countries and across systems, under the assumption of independence of these two variables. A value of 1.0 indicates that the number of institutions from C in R is 'as expected'. See the legend to Table 4 for an exact definition. Table 4 gives for each ranking system the five most 'preferred' countries. It reveals differences in geographical coverage among ranking systems. It shows the orientation of ***U-Multirank*** towards Europe, ***ARWU*** towards North America and Western Europe, LEIDEN towards emerging Asian countries and North America, and ***QS*** and ***THE*** towards Anglo-Saxon countries, as Great Britain, Canada and Australia appear on both.

Table 4. Five most 'preferred' countries per ranking system

| System | Country | Nr. Univs | Preference | System | Country | Nr. Univs | Preference |
|---|---|---|---|---|---|---|---|
| ARWU | Canada | 20 | 2.1 | THE | Taiwan | 24 | 2.0 |
| | USA | 146 | 2.1 | | Great Britain | 78 | 1.9 |
| | Netherlands | 12 | 2.1 | | Australia | 31 | 1.8 |
| | Great Britain | 20 | 1.9 | | Canada | 25 | 1.7 |
| | Germany | 39 | 1.5 | | Japan | 41 | 1.4 |
| LEIDEN | China | 108 | 1.9 | U-MULTI-RANK | Netherlands | 20 | 1.3 |
| | Korea | 33 | 1.8 | | Spain | 67 | 1.3 |
| | Canada | 28 | 1.8 | | Poland | 45 | 1.3 |
| | Taiwan | 19 | 1.5 | | Germany | 84 | 1.3 |
| | USA | 173 | 1.5 | | Portugal | 27 | 1.3 |

| | | | | |
|---|---|---|---|---|
| | Australia | 33 | 1.7 | |
| | Great Britain | 75 | 1.6 | |
| QS | Brazil | 22 | 1.6 | |
| | Canada | 26 | 1.5 | |
| | Korea | 27 | 1.4 | |

Legend to Table 4. The preference P of ranking system R for a particular country C is defined as follows. If $n[i,j]$ indicates the number of institutions from country i in system j, $\sum_i n[i,j]$ the sum of $n[i,j]$ over all i (countries), and $\sum_j n[i,j]$ the sum of $n[i,j]$ over all j (systems), $P = (n[i,j] / \sum_i n[i,j]) / (\sum_j n[i,j] / \sum_i \sum_j n[i,j])$.

A second way to analyse differences in geographic orientation among ranking systems focuses on the *top 100 lists* in the **ARWU**, **QS** and **THE** rankings based on their overall score and on the two **Leiden top lists**, rather than on the total set of covered institutions analysed in Table 4, and identifies for each system the country of location of 'unique' institutions, i.e., universities that appear in a system's the top list but that are not included in the top list of any other system. The results presented in Table 5 are not fully consistent with those in Table 4, due to differences among countries in the frequency at which their institutions appear in top 100 lists, but there is a considerable agreement between the two tables. Table 5 reveals that in the **ARWU** and the **Leiden CIT** top list most unique institutions are from the USA, and in the **QS** top from Great Britain and two Asian entities: Korea and Hong Kong (formally a part of China). Unique institutions in the **Leiden PUB** top list are especially located in China, and, to a letter extent, in Italy, and those in the **THE** top list in Germany, USA and The Netherlands.

Table 5. Country of location of unique institutions in top 100 lists

| Ranking system | Nr unique univs | Country of location with >=2 univs |
|---|---|---|
| ARWU | 11 | USA (4), Israel (2) |
| THE | 8 | Germany (3), USA (2), Netherlands (2) |
| QS | 14 | Great Britain (3), Hong Kong (2) Korea (2) |
| LEIDEN-PUB | 11 | China (6), Italy (2) |
| LEIDEN-CIT | 26 | USA (9), Great Britain (6), Switzerland (2) France (2) |

## 4. Indicator scores and their distributions

*Missing values*

In the **ARWU**, **THE** and **QS** rankings the *overall* indicators are presented only for the first 100, 200 and 400 universities, respectively. In addition, **QS** presents on its website for *all* its indicators only values for the first 400 institutions. Occasionally, values are missing. This is true, for instance, in the **QS** system for the values of Rockefeller University on the indicators Academic Reputation, Employer Reputation and Overall Score. As regards **U-Multirank**, not all universities have participated in the surveys per subject field, and those who did were not necessarily involved in each subject field. Of the about 1,300 institutions retrieved from the **U-Multirank** website, 28 per cent has a score for the indicator quality of teaching in at least one subject field, and 12 per cent in at least three fields.

*From data to indicators*

Both **ARWU** and **QS** apply the method of *normalizing by the maximum*: for each indicator, the highest scoring institution is assigned a score of 100, and other institutions are calculated as a percentage of the top score. Standard statistical techniques are used to adjust the indicator if necessary. The **QS** documentation adds that for some indicators a cut-off is applied so that multiple institutions have score 100. In fact, for the indicators citations per faculty, academic reputation and employer reputation the number of institutions with score 100 is 10, 12 and 11, respectively.

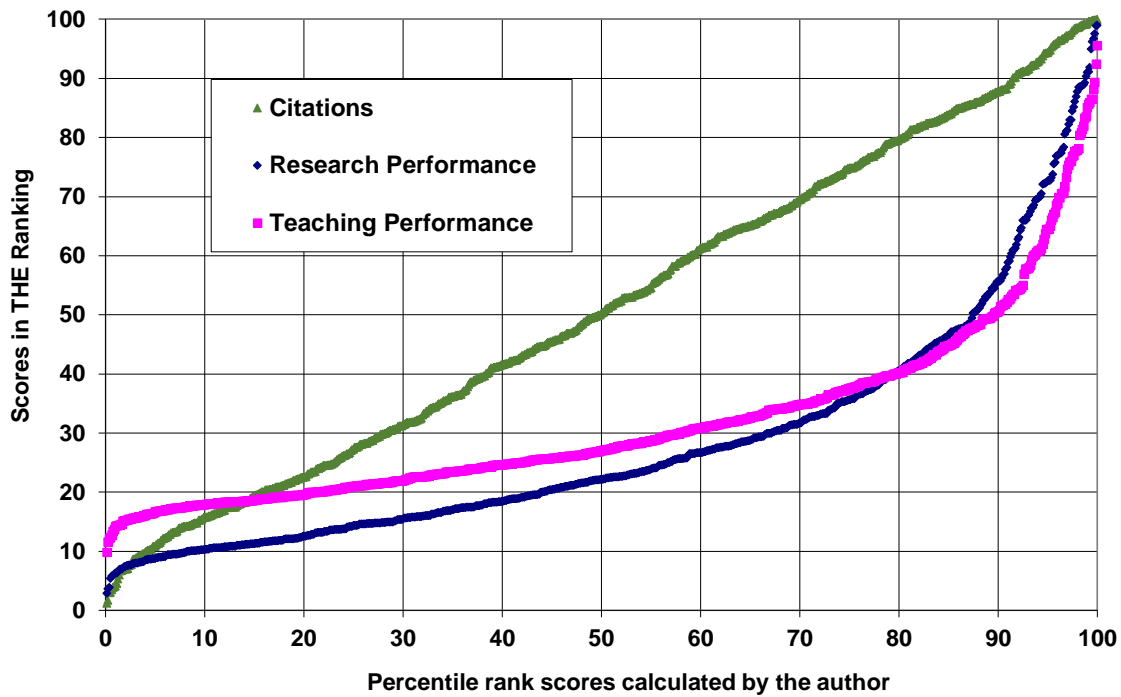

Figure 1. Scores of three key indicators in the **THE** ranking plotted against their percentile rank calculated by the author of the current article

The **THE** system applies a *percentile rank-based approach*: For all indicators except the Academic Reputation Survey, a cumulative probability function is calculated, and it is evaluated where a particular institution's indicator sits within that function, using a version of Z-scoring. For the Academic survey, an exponential component is added. This is illustrated in Figure 1. It plots the scores in the **THE** Ranking 2016 against percentile rank scores calculated by the author of this article. For the citations all observations are plotted on the diagonal. This illustrates that **THE** citation scores are in fact percentile rank scores. Figure 1 reveals how radically the **THE** research and teaching performance scores deviate from percentile rank scores, and how strong the exponential component is. 90 per cent of institutions has a Research or Teaching Performance Score below 55 or 50, respectively.

**U-Multirank** applies a '*distance to the median*' approach. Per indicator, universities are assigned to 5 performance groups ranging from excellent (=A) to weak (=E), based on the distance of the score of an individual institution to the median performance of all institutions that **U-Multirank** has data for. It should be noted that the distribution of indicator values (A-E) may substantially vary from one indicator to another, and deviates strongly from a distribution based on quintiles. For instance, as regards the absolute number of publications the percentage of institutions with score A, B, C, D and E is 2.6, 47.3, 25.5, 20.7 and 0.0, respectively (for 3.9 % no value is available). For the number of publications cited in patents these percentages are 30.6, 7.4, 11.6, 30.3 and 8.8 (for 11.2 % no value is available), and for the number of post doc positions 15.3, 4.0, 3.9, 15.3 and 5.0 (for 56.5 % data is unavailable).

*Skewness of indicator distributions*

Table 6 presents for a group of 17 indicators the skewness of the indicator distributions related to all institutions for which data are available. Figure 2 visualizes the distribution of 7 key indicators by plotting the institutions' scores as a function of their rank. Table 5 shows that the **Leiden** absolute number of 'top' publications, - i.e., the number of publications among the 10 per cent most frequently cited articles published worldwide – has the highest skewness, and the **THE** citations indicator the lowest. The latter result is not surprising, as Figure 1 revealed already that the values obtained by this indicator are percentile ranks, for which the skewness is mathematically zero. Disregarding **Leiden** Number of Top Publications and **THE** Citations, the 5 **ARWU** indicators have the highest skewness, followed by 3 **THE** indicators, and 4 **QS** jointly with the two **Leiden** relative impact indicators the lowest.

Table 6. Skewness of 17 indicator distributions

|  | All Universities | |
| --- | --- | --- |
|  | Nr. Univs | Skew-ness |
| LEIDEN Nr. Top Publications (Top 10%) | 840 | 4.03 |
| ARWU Awards | 500 | 3.03 |
| LEIDEN Publications | 840 | 2.56 |
| ARWU Alumni | 500 | 2.55 |
| ARWU Publ in Nature, Science | 498 | 2.30 |
| ARWU World Rank | 100 | 2.08 |
| ARWU Highly Cited Researchers | 500 | 1.81 |
| THE Teaching | 799 | 1.63 |
| THE Research | 799 | 1.49 |
| THE Overall | 199 | 1.01 |
| QS Overall | 400 | 0.65 |
| LEIDEN % Top Publications (Top 10%) | 840 | 0.54 |
| LEIDEN Mean Normalized Citation Score (MNCS) | 840 | 0.46 |
| QS Academic Reputation | 400 | 0.43 |
| QS Employer Reputation | 400 | 0.36 |
| QS Citations per Faculty | 399 | 0.26 |
| THE Citations | 799 | 0.07 |

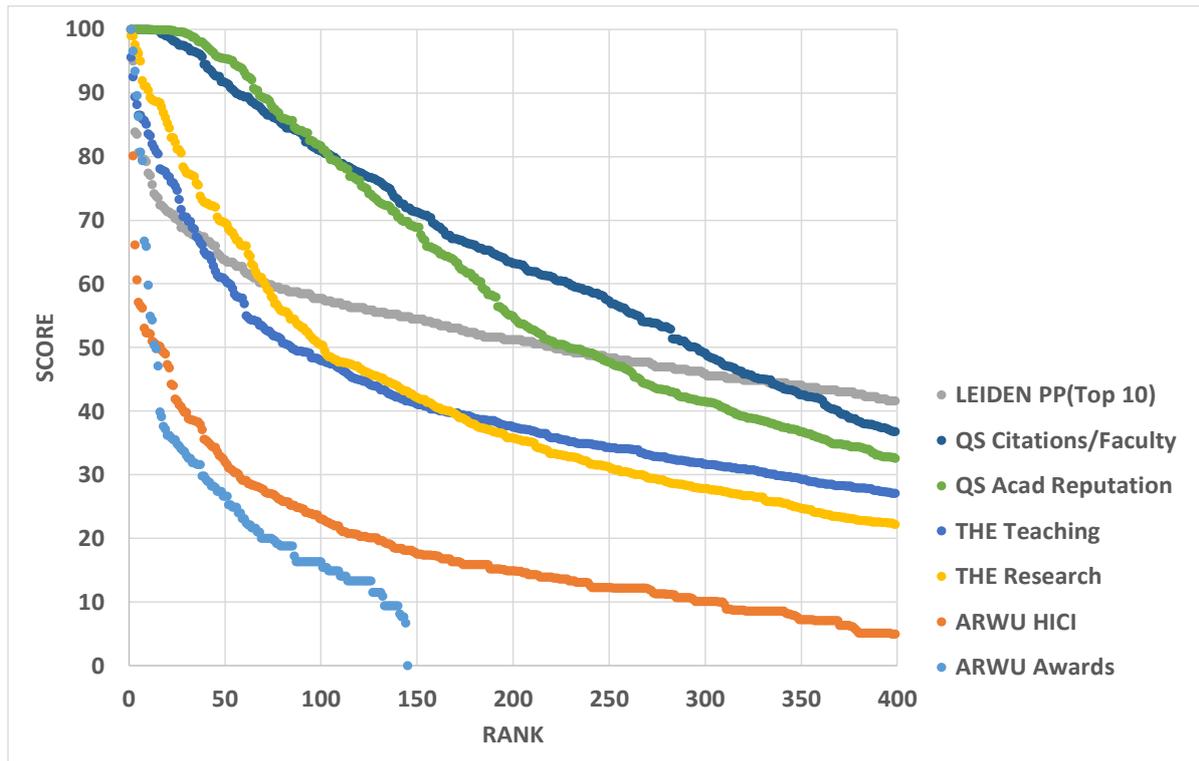

Figure 2. institutions' scores as a function of their ranks in 7 key indicator distributions. Legend: LEIDEN PP (Top 10): The percentage of publications among the top 10 per cent most frequently cited articles published worldwide. THE Teaching: THE Teaching Performance; THE Research: THE Research Performance. ARWU HICI: ARWU Highly Cited Researchers.

## 5. Statistical correlations

Tables 7-9 presents the Spearman coefficients (denoted as Rho) of the rank correlation between pairs of selected indicators, arranged into 3 groups: a group with pairs of seemingly identical indicators related to staff, student and funding data; citation-based indicators; and a group combining reputation- and recognition-based indicators with key indicators from the group of the citation-based measures. The correlations between two indicators are calculated for those institutions that have non-missing values for both measures. Row N gives the number of institutions involved in a calculation. Unless indicated differently, all correlations in Tables 7-9 are statistically significant at the p=0.001 level. Rank correlations above 0.8 are printed in bold, and those below 0.4 in bold and italic. If one qualifies correlations with absolute values in the range 0.0-0.2, 0.2-0.4, 0.4-0.6, 0.6-0.8 and 0.8-1.0 as 'very weak', 'weak', 'moderate', 'strong' and 'very strong', respectively, it can be said that correlations printed in bold-but-not-italic are very strong; correlations in bold and italic are weak or very weak, while all other are moderate or strong.

Unsurprisingly, a very strong correlation is found between an institution's number of publications in the **ARWU** ranking and that in the **Leiden** Ranking (Rho=0.96, n=468), as both numbers are extracted from the Web of Science. On the other hand, the **ARWU** number of publications in Nature and Science correlates 0.73 with the **Leiden** (absolute) number of 'top' publications, suggesting that top publications are not merely published in these two journals.

Table 7: Spearman rank correlations between specific pairs of identical/very similar variables from different sources

| Variable 1 | Variable 2 | Statistic | Score |
|---|---|---|---|
| QS Internat. students | THE % internat. students | Rho | **0.87** |
| | | | |
| | | N | 311 |
| QS Faculty-Student Ratio | THE Student-Staff Ratio | Rho | - 0.47 |
| | | N | 289 |
| QS Internat. Faculty | UMULTI Internat. Acad. Staff | Rho | *0.13* |
| | | N | 107 |
| THE Industry Income | UMULTI Income from private sources | Rho | 0.48 |
| | | N | 201 |

The most striking outcome in Table 7 is that the **QS** Faculty-Student Ratio correlates only moderately with the **THE** student-staff ratio (rho=-0.47). From the data descriptions in the two systems it does not become clear why there are such large differences between the two. This is also true for the very weak correlation between **QS** International Faculty and **U-Multirank**'s International Academic Staff.

Table 8: Spearman rank correlations between citation-based indicators

| | | LEIDEN MNCS | LEIDEN % Publ. in Top 10% | QS Citation per Faculty | THE Citations | UMULTI Top Cited Publ |
|---|---|---|---|---|---|---|
| ARWU Highly Cited Researchers | Rho | 0.69 | 0.70 | *0.38* | 0.70 | 0.61 |
| | N | 468 | 468 | 308 | 416 | 461 |
| LEIDEN MNCS (Mean Normalized Citation Rate) | Rho | | **0.98** | *0.32* | **0.92** | **0.86** |
| | N | | 840 | 344 | 589 | 742 |
| LEIDEN % Publ. in Top 10% Most Cited Articles | Rho | | | *0.34* | **0.92** | **0.89** |
| | N | | | 344 | 589 | 742 |
| QS Citation per Faculty | Rho | | | | *0.38* | *0.26* |
| | N | | | | 348 | 343 |
| THE Citations | Rho | | | | | **0.81** |
| | N | | | | | 620 |

Noteworthy in Table 8 is first of all the very high correlation between the two **Leiden** citation impact measures (rho=0.98). Apparently, at the level of institutions it does not make a difference whether one focuses on the mean (MNCS) or the top of the citation distribution. Interestingly, also the **THE** Citation indicator shows a strong correlation with the **Leiden** impact measures. The description of this measure on the **THE** Ranking Methodology page (**THE** Ranking Methodology, n.d.) suggests that it is most similar if not identical to the **Leiden** MNCS, but a key difference is that it is based on Scopus, while the **Leiden** indicators are derived from the Web of Science. The **U-Multirank** indicator of top cited publications is provided by the Leiden Centre for Science and Technology Studies using the same methodology as that applied in the **Leiden** Ranking. The most remarkable outcome in Table 7 is perhaps that the indicator **QS** Citation per Faculty shows only a weak correlation with the other citation-based indicators. This result is further analysed in Section 6 below.

Table 9. Spearman correlations between citation, reputation and teaching-related indicators

| | | ARWU Highly Cited Res | LEIDEN MNCS | THE Research | THE Teaching | QS Acad Reput | QS Citations/ Faculty | UMULTI Quality Teaching |
|---|---|---|---|---|---|---|---|---|
| ARWU Awards | Rho | 0.43 | 0.50 | 0.46 | 0.47 | 0.45 | *0.30* | *0.09\** |
| | N | 500 | 468 | 416 | 416 | 314 | 308 | 60 |
| ARWU Highly Cited Res | Rho | | 0.69 | 0.64 | 0.60 | 0.53 | *0.38* | *0.22\** |
| | N | | 468 | 416 | 416 | 314 | 308 | 60 |
| LEIDEN Mean Norm.Citation Score (MNCS) | Rho | | | 0.60 | 0.54 | 0.41 | *0.32* | *0.36* |
| | N | | | 589 | 589 | 349 | 344 | 82 |
| THE Research | Rho | | | | **0.81** | 0.76 | 0.52 | 0.42 |
| | N | | | | 799 | 356 | 348 | 94 |
| THE Teaching | Rho | | | | | 0.76 | 0.50 | 0.43 |
| | N | | | | | 356 | 348 | 94 |
| QS Aademic Reputation | Rho | | | | | | *0.34* | *0.33\*.* |
| | N | | | | | | 264 | 53 |
| QS Citations per Faculty | Rho | | | | | | | 0.47 |
| | N | | | | | | | 29 |

* Not significant at p=0.05.

Table 9 presents pairwise correlation coefficients between seven citation-, reputation- or teaching-related indicators. The only very strong rank correlation is that between **THE** Research and **THE** Teaching. Both measures are composite indicators in which the outcomes of a reputation survey constitute the major component. On the **THE** Ranking Methodology page it is unclear whether the reputation components in the two indicators are different. The very strong correlation between the two indicators seems to suggest that these components are very similar if not identical.

The weak correlation between **QS** Citations per Faculty and other citation-based indicators has already been mentioned above. Table 9 shows that there is also a weak rank correlation inside the **QS** system between the citation and the academic reputation measure (Rho=0.34). The major part of the pairs shows moderate or strong, positive Spearman correlation coefficients.

The **U-Multirank** Quality of Teaching score in Table 9 is calculated by the author of the current paper, based on the outcomes of the survey among students, conducted by the **U-Multirank** team in 13 selected subject fields, and mentioned in Section 1. For institutions participating in at least two surveys, the performance classes (A-E) were quantified (A=5, B=4, etc.), and an average score was calculated over the subject fields. The number of cases involved in the calculation of the rank correlation coefficients between this indicator and other measures is relatively low, and the major part of the coefficients are not statistically significant at p=0.05.

## 6. Secondary analyses

### 6.1 Characteristics of national academic systems

A secondary analysis based on *THE* data examined for the 19 major countries with more than 10 institutions the rank correlation between *THE* Citations and *THE* Research Performance. According to the *THE* Ranking Methodology Page, the citation-based (research influence) indicator is defined as the number of times a university's published work is cited by scholars globally, compared with the number of citations a publication of similar type and subject is expected to have. *THE* Research Performance is a composite indicators based on three components: Outcomes of a Reputation Survey (weight(W)=0.6); Research income (W=0.2); and Research productivity (W=0.2).

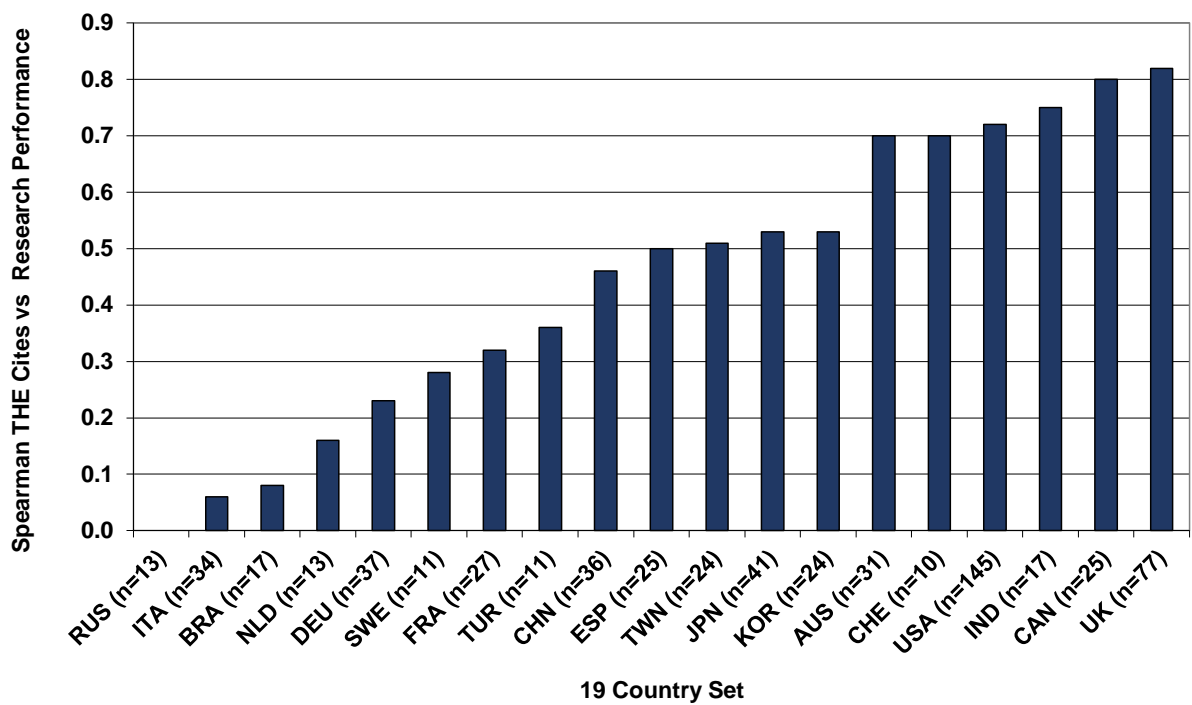

Figure 3. Spearman rank correlation coefficient between Citations and Research Performance per country (Data source: *THE* Ranking 2016)

The results are presented in Figure 3. Countries can be categorized into three groups. A first group with rho scores up or above 0.7 consists of four Anglo-Saxon countries, India and Switzerland. A second group, with scores between 0.4 and 0.6 contains four Asian countries and Spain. Finally, the group with scores below 0.4 includes four Western-European countries, Turkey and Russia, and also Brazil. As an illustration, Figures 4 and 5 present a scatterplot representing the scores of the institutions in Italy and The Netherlands, respectively. In Italy, but also in Brazil and Russia, a large subset of universities has statistically similar Research Performance scores, but assumes a wide range of citation scores; at the same time, a few universities with high Research Performance scores have median or low citation scores. The Netherlands and Germany show a different, partly opposite pattern: a relatively large set of universities has similar, high citations scores, but reveals a wide range of Research Performance scores. Both patterns result in low rank correlation coefficients.

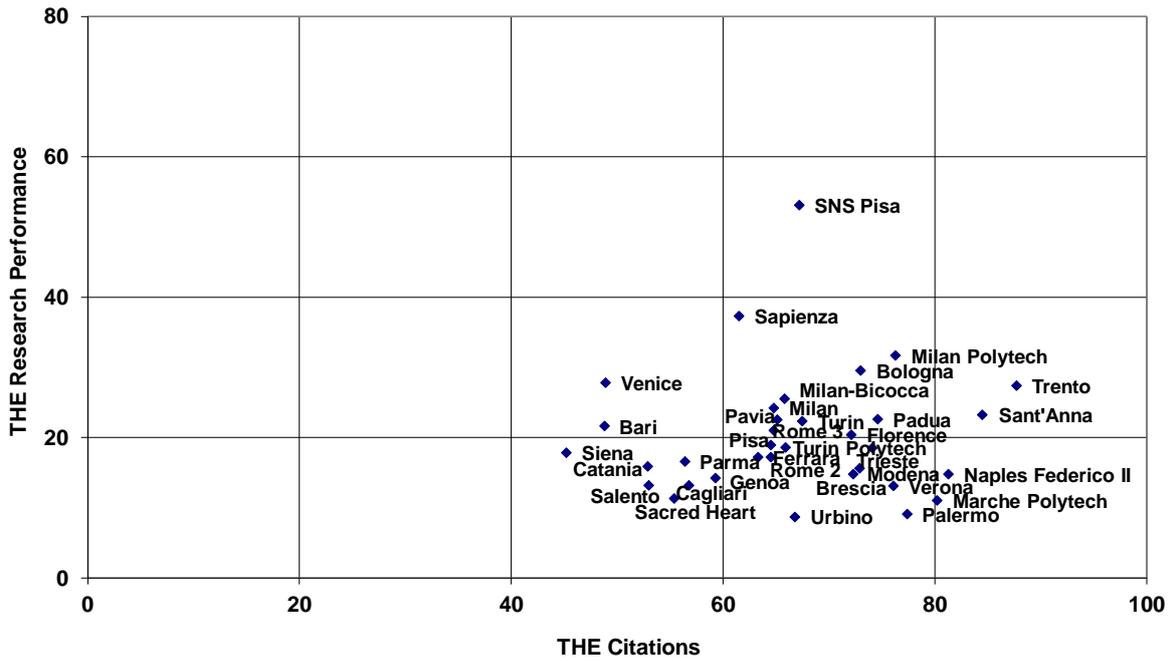

Figure 4. Scatterplot of THE Research Performance vs. THE Citations for Italy

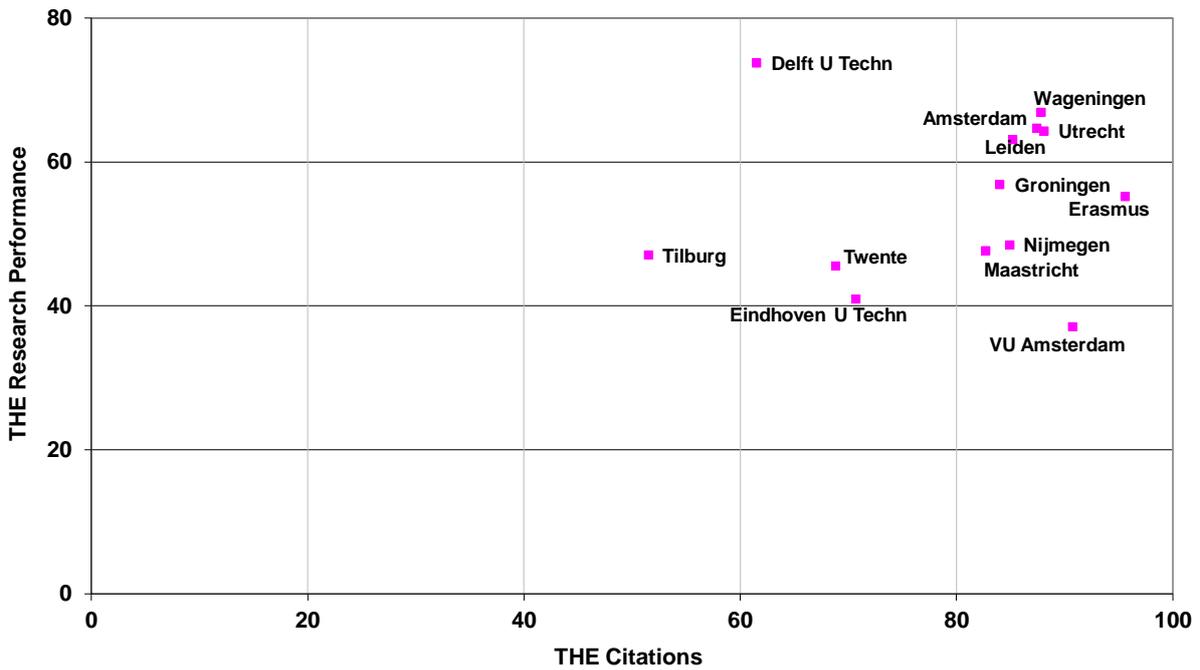

Figure 5. Scatterplot of THE Research Performance vs. THE Citations for The Netherlands

The interpretation of the observed patterns is unclear. The figure suggests that there are differences among global geographical regions. A low correlation may reflect a certain degree of conservatism in the national academic system in the sense that academic reputation is based on performances from a distant past, and does not keep pace well enough with recent performances as reflected in citations.

*6.2 QS versus Leiden citation-based indicators*

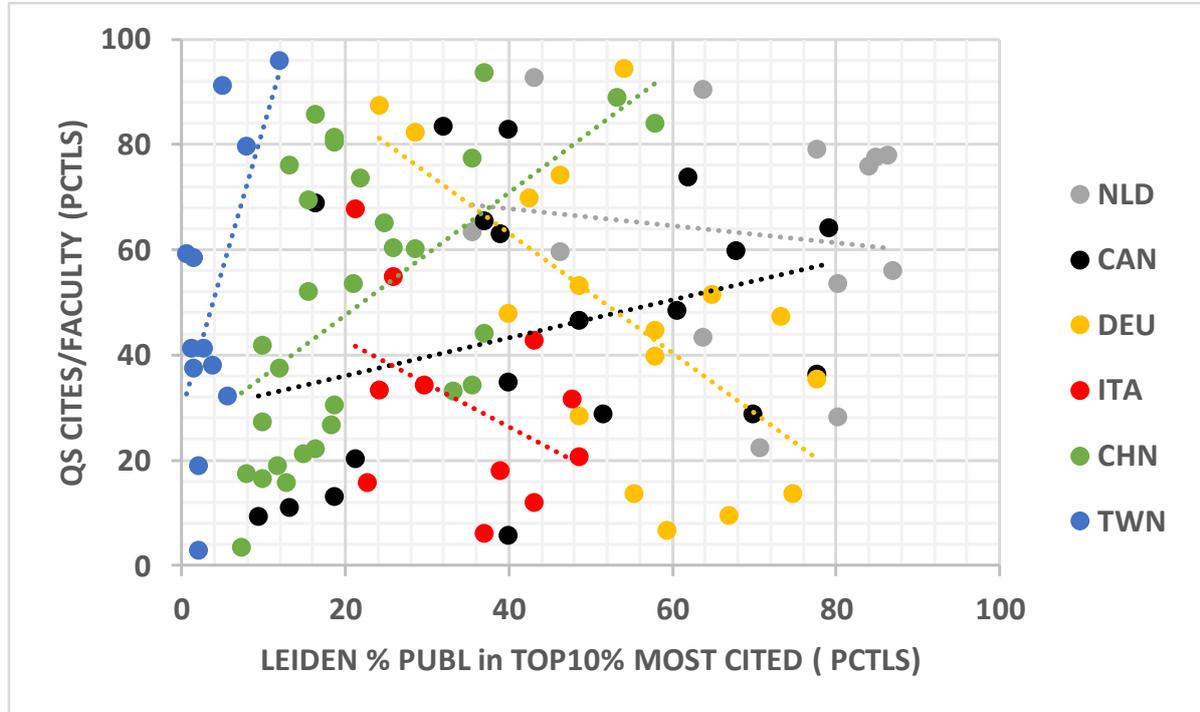

Figure 6. QS and Leiden citation impact indicators for institutions in 6 selected countries

Figure 6 plots for institutions in 6 countries the scores on the **QS** Citations Per Faculty indicator against the **Leiden** percentage of publications among the top 10 per cent most frequently cited documents published worldwide. Both scores were expressed as percentile ranks by the current author. For details on the QS measure the reader is referred to QS Normalization (n.d.) and QS Methodology (n.d.) and on the **Leiden** indicators to Leiden Indicators (n.d.).

Five countries in Figure 6 have institutions among the top 20 per cent worldwide in the QS ranking, seemingly regardless of their citation scores on the **Leiden** indicator: Taiwan, Germany and The Netherlands have three institutions, China (including Hong Kong) six, and Canada two. This outcome raises the question whether the **QS** measure applies 'regional weightings' to correct for differences *in citation counts* between world regions, analogously to the application of regional weightings to counter discrepancies in response rates in the **QS** *Academic Reputation survey*. It must be noted that the current author could not find an explicit reference to such weightings in the **QS** document on normalization (QS Normalization, n.d), although this document does indicate the use of weightings by scientific-scholarly discipline.

A second normalization of the **QS** measure calculates the ratio of citations and number of faculty. Interestingly, this leads to a negative correlation with the **Leiden** measure for Italy, The Netherlands, and, especially, for Germany, two institutions in which – *Humboldt University Berlin* and *University of Heidelberg* – have a **Leiden** percentile rank above 60 but a **QS** Citation per Faculty percentile rank below 20.

*6.3 THE Research Performance versus QS Academic Reputation*

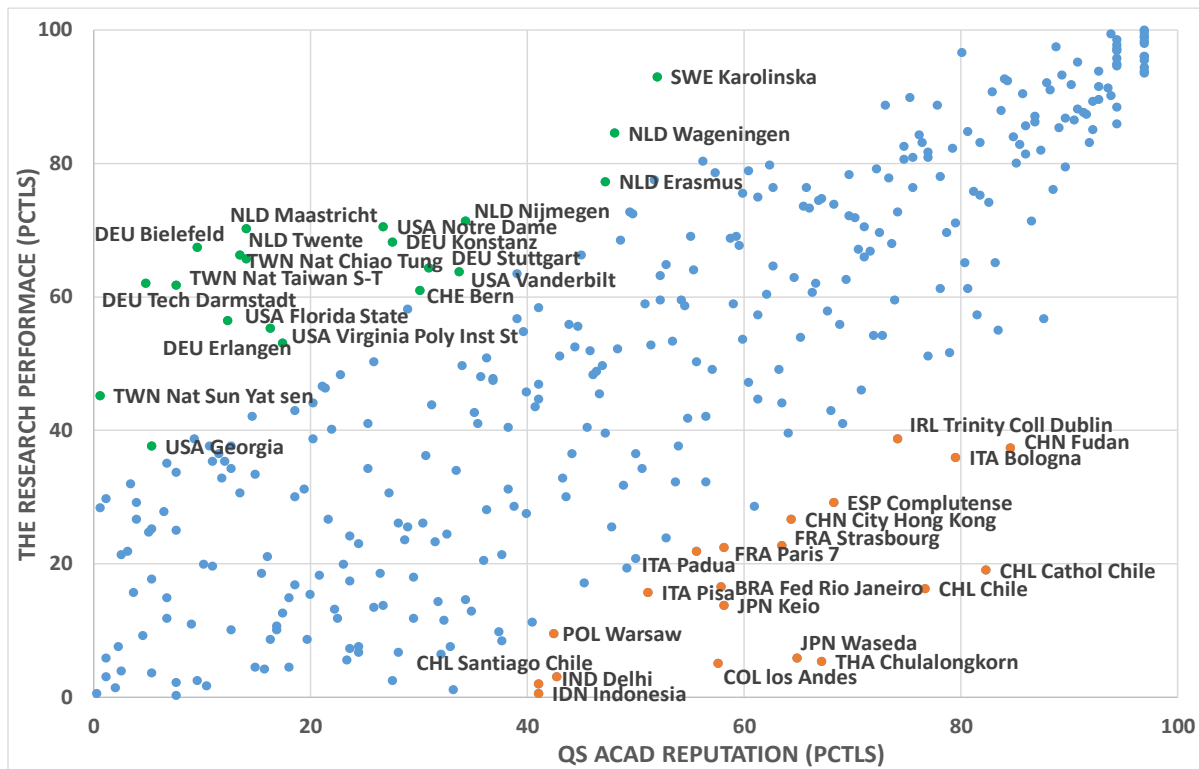

Figure 7. Scatterplot of THE Research Performance versus QS Academic Reputation

Figure 7. presents a scatterplot of the reputation-based **THE** Research Performance against **QS** Academic Reputation. As in the previous secondary analysis in Section 6.1, both measures were expressed as percentile ranks by the current author. The figure displays the names of the top 20 institutions with the largest, and the bottom 20 with the smallest difference between the **THE** and the **QS** measure, respectively. Focusing on countries appearing at least twice in a set, institutions in the top 20 set, for which the **THE** score is much larger than the **QS** score, are located in The Netherlands, Germany, USA and Taiwan, while universities in the bottom 20 set can be found in Chili, Italy, France and Japan.

These differences are probably caused by the fact that in the **QS** methodology 'regional weightings are applied to counter any discrepancies in response rates' (QS Normalization, n.d.), while **THE** does not apply such weighting. Hence, in the top 20 set one finds institutions from countries that have already a sufficient number of institutions in the upper part of the reputation ranking, and in the bottom 20 set universities in countries that are underrepresented in this segment. The outcomes then would suggest that Southern Europe and Northern Europe are considered distinct regions in the QS approach.

*6.4 ARWU Highly Cited Researchers vs. Leiden Top Publications indicator*

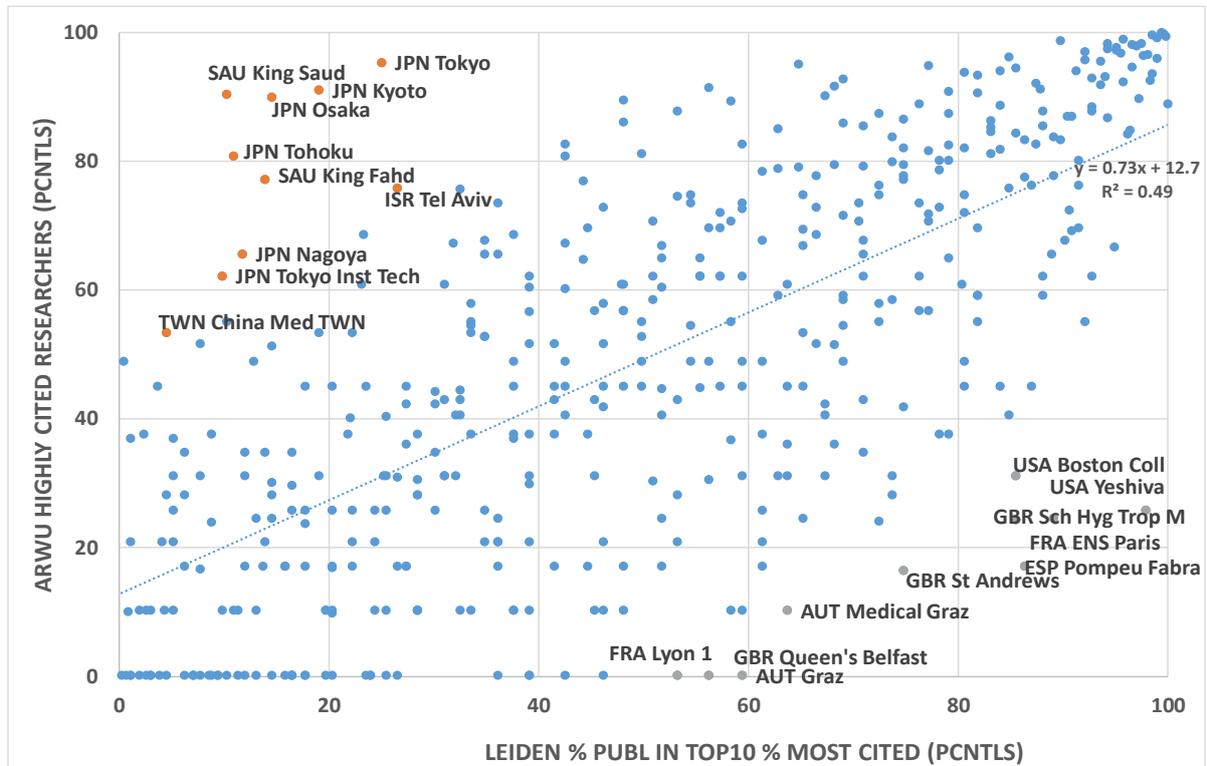

Figure 8. Scatterplot of ARWU Highly Cited Researchers versus Leiden Top Publications indicator

Figure 8 is constructed in a manner very similar to Figure 7, but for two different indicators. It gives the names of the top 10 institutions with the largest difference, and the bottom 10 with the smallest difference between **ARWU** and **Leiden** measure. In the top 10 set two institutions from Saudi Arabia appear. Their score on the Highly Cited Researchers linked with these institutions indicator is much higher than 'expected' on the basis of the number of highly cited articles published from them.

This outcome illustrates a factor highlighted by Gingras (2014) who found in the Thomson Reuters List of Highly Cited Researchers – the data source of the **ARWU** indicator – a disproportionally large number of researchers linked with institutions in Saudi Arabia, mostly via their secondary affiliations, and who suggested that "by providing data on secondary affiliations, the list inadvertently confirms the traffic in institutional affiliations used to boost institutions' places in world university rankings". King Abdulaziz University, the institution Gingras found to be the most 'attractive' given the large number of researchers that indicated its name as secondary affiliation, is not in the Top 20 list, but it ranks 28th and would have been included in a top 30 list. The top 10 list includes six Japanese institutions. Whether their score on the **ARWU** Highly Cited Researchers indicator is caused by the same factor is as of yet unclear, and needs further investigation, without which *no* valid conclusions about these institutions can be drawn.

The institutions and countries represented in the bottom 10 set seem to constitute prima facie a rather heterogeneous set. However, it includes a number of institutions focusing on social sciences, or located in non-English speaking countries. This suggests that the **Leiden** indicator corrects more properly for differences between subject fields and native languages than the TR List of Highly Cited Researchers does.

It must be noted that the **ARWU** indicator is based on two lists of highly cited researchers, both compiled by Thomson Reuters, a first one in 2001, and a new one in 2013. The **ARWU** 2015 ranking is based on the sum of the numbers in the two lists. But the counts derived from the new list are based exclusively on the *primary* affiliation of the authors, thus substantially reducing the effect of secondary affiliations highlighted by Gingras.

## 7. Discussion and conclusions

The *overlap* analysis clearly illustrates that there is *no* such set as '*the*' top 100 universities in terms of excellence: it depends on the ranking system one uses which universities constitute the top 100. Only 35 institutions appear in the top 100 lists of all 5 systems, and the number of overlapping institutions per pair of systems ranges between 49 and 75. An implication is that national governments executing a science policy aimed to increase the number of academic institutions in the 'top' of the ranking of world universities, should not only indicate the range of the top segment (e.g., the top 100), but also specify which ranking(s) are used as a standard, and argue why these were selected from the wider pool of candidate world university rankings.

Although most systems claim to produce rankings of *world* universities, the analysis of *geographical coverage* reveals substantial differences between the systems as regards the distribution of covered institutions among geographical regions. It follows that the systems define the 'world' in different manners, and that – compared to the joint distribution of the 5 systems combined – each system has a proper orientation or bias, namely **U-Multirank** towards Europe, **ARWU** towards North America, **Leiden Ranking** towards emerging Asian countries, and **QS** and **THE** towards Anglo-Saxon countries.

*Four* entirely different methods were applied to construct *indicator scores* from raw data. **ARWU** and **QS** apply a normalization by the maximum, **THE** uses a percentile rank-based approach but for some indicators an exponential component was added, while **U-Multirank** calculates a distance to the median. This has severe implications for the *interpretation* of the scores. For instance, in the **THE** system 90 per cent of institutions has a Research or Teaching Performance score below 55 or 50, respectively. This means that only a small fraction of institutions 'profits' in the overall ranking from a high score of these indicators, reflecting that the distribution of the actual values of the reputation-based component is much more skewed than that for the citation-based indicator. The distribution of **U-Multirank** performance classes (A-E) among institutions varies substantially between indicators, and, as the definition of the classes is based on the distance to the median rather than on quintiles of a distribution, may strongly deviate from 20 per cent.

**ARWU** indicators (Awards, Alumni, Articles in Nature and Science, Highly Cited Researchers, and Overall) show the largest *skewness* in their distributions, followed by **THE** indicators (Research and Teaching Performance, Overall), while **QS** indicators (Academic and Employer Reputation and Overall) jointly with the two **Leiden** relative citation impact indicators obtain the lowest skewness values. It follows that the degree of skewness measured in the various systems is substantially affected by the way in which the systems calculate the indicator scores from the raw data.

Several pairs of very similar if not identical indicators from *different* ranking systems rank-correlate only *moderately*, especially those based on student and faculty numbers. The causes of this lack of correlation are as yet unclear and must be clarified. It must be noted that in several systems the role of this type of data is far from being marginal. For instance, in the QS citation impact indicator an institution's number of academic staff constitutes the denominator in a citation-per-faculty ratio for that institution. Also, the question should be addressed whether *self-reported* data from institutions are sufficiently accurate to constitute an important factor in the calculation of indicators and rank positions. But even if data is obtained from statistical agencies such as national statistical offices, a thorough investigation is urgently needed as to whether such agencies apply the same definitions and categorizations in the data collection and reporting.

The citation-based indicators from **Leiden**, **THE**, **ARWU** and **U-Multirank** show strong or very strong rank correlations with one another, but correlate only weakly with the **QS** Citation per Faculty indicator. The latter is constructed differently in that an institution's total citation count, corrected for differences in citation levels between disciplines, is divided by the number of faculty employed in an institution. An analysis comparing **QS** and **Leiden** citation indicator scores may suggest that the **QS** citation measure does not only apply a *field* normalization, but also a normalization by *geographical region*, but more research is needed to validate this. The effect of indicator normalization is further discussed below.

A pairwise *correlation* analysis between seven citation-, reputation- or teaching-related indicators from the 5 systems shows for the major part of the pairs moderate or strong – but never very strong –, positive Spearman correlation coefficients (with values between 0.4 and 0.8). The conclusion is that these indicators are related to one another, but that at the same time a certain degree of *complementarity* exists among the various ranking systems, and that the degree of (dis-)similarity between indicators *within* a ranking system is similar to that between measures from *different* systems. The conclusion is that the various ranking methodologies do indeed measure different aspects. There is no single, 'final' or 'perfect' operationalization of academic excellence.

The analysis on the statistical relation between *two reputation-based indicators*, namely the **QS** Academic Reputation indicator, and the **THE** Research Performance measure, which is largely based on the outcomes of the **THE** reputation survey, reveals the effect of the use of 'weightings' to counter discrepancies or unbalances upon the overall results. This particular case relates to (world) regional weightings. A ranking seems to naturally direct the attention of users to its top, and multiple rankings to multiple tops. But what appears in the top very much depends upon which normalizations are carried out.

This analysis, as well as the analysis of the QS citation-per-faculty measure discussed above, provides an illustration of how the position of institutions in a ranking can be influenced by using proper, effective indicator normalizations. The current author does not wish to suggest that the developers intentionally added a normalization to boost particular sets of institutions or countries, as they provide in their methodological descriptions purely methodological considerations (QS Normalization, n.d.). But the two analyses clearly show how such targeted, effective boosting *could in principle* be achieved technically. When ranking systems calculate complex, weighted or normalized indicators – as they often do –, they should at the same time provide simple tools to show users the actual *effect* of their weightings or normalizations. Figures 7 and 8 in Section 6 illustrate how this could be done.

The analysis focusing on the number of highly cited researchers reveals possible traces of the effect of 'secondary' affiliations of authors in counting the number of highly cited researchers per institution. The **ARWU** team has already adjusted its methodology to counter this effect. But even if secondary affiliations are fully ignored, this indicator can be problematic in the assessment of an institution. How should one allocate (highly cited) researchers to institutions as researchers move from one institution to another – a notion that is properly expressed in the methodology along which **ARWU** calculates its Awards and the Alumni indicator. The analysis has identified other universities showing discrepancies similar to those of Saudi institutions, but the interpretation of this finding is as yet unclear. A general conclusion holds that by systematically comparing pairs of indicators within or across systems, discrepancies may be detected that ask for further study, and help evaluating the data quality and validity of indicators.

The analysis on the correlation between *academic reputation and citation impact* in the ***THE*** ranking (see Figures 4 and 5 in Section 6) shows first of all that two-dimensional scatterplots for a subset of institutions with labelled data points provide a much more comprehensive view of the relative position of individual institutions than the view one obtains by scanning one or more rank lists sequentially from top to bottom. The outcomes of the analysis raise interesting questions. Why are there such large differences between countries as regards the correlation between the two types of indicators? What does it mean if one finds for a particular country that a large subset of institutions has statistically similar citation impact scores, but assumes a wide range of reputation-based scores, or vice versa?

The current author wishes to defend the position that ranking systems would be more useful if they would raise this type of questions, enable users to view the available empirical data that shed light on these questions, and in this way contribute to their knowledge on the pros and cons of the various types of indicators, rather than to scan sequentially through different rankings, or calculate composite indicators assigning weights to each constituent measure.

*Concluding remarks*

Developers of world university ranking systems have made enormous progress during the past decade. Their systems are currently much more informative and user friendly than they were some 10 years ago. They do present a series of indicators, and institutions ran be ranked by each of these separately. But the current interfaces seem to hinder a user to obtain a comprehensive view. It is like looking into the outside world through a few vertical splits in a fence, one at the time. In this sense, these systems are still one-dimensional. A system should not merely present a series of separate rankings in parallel, but rather a dataset and tools to observe patterns in multi-faceted data. The simple two dimensional scatterplots – to which easily a third dimension can be added by varying the shape of the data point markers – are good examples.

Through the selection of institutions covered, the definition of how to derive ratings from raw data, the choice of indicators and the application of normalization or weighting methodologies, a ranking system distinguishes itself from other rankings. Each system has its proper orientation or 'profile', and there is no 'perfect' system. To enhance the level of understanding and adequacy of interpretation of a system's outcomes, more insight is to be provided to users into the differences between the various systems, especially on how their orientations influence the ranking positions of given institutions. The current paper has made a contribution to such insight.

## Acknowledgements

The author wishes to thank two referees for their useful comments on an earlier version of this paper. The author is also grateful to the members of the Nucleo di Valutazione of the Sapienza University of Rome for stimulating discussions about the interpretation and the policy significance of world university rankings.

**Table A1: Overview of five information systems on the performance of higher education institutions**

| Aspect | ARWU World University Rankings 2015 | CWTS Leiden Ranking 2016 | QS World University Rankings 2015-2016 | THE World University Rankings 2015-2016 | U-Multirank 2016 Edition |
|---|---|---|---|---|---|
| Website | http://www.shanghairanking.com/ARWU2015.html | http://www.leidenranking.com/ | http://www.topuniversities.com/university-rankings | https://www.timeshighereducation.com/world-university-rankings | http://www.umultirank.org |
| Universities included | Every university that has any Nobel Laureates, Fields Medallists, Highly Cited Researchers, or papers published in Nature or Science, or significant amount of papers indexed by SCIE/SSCI. The best 500 are published on the web. | All 842 universities worldwide with more than 1000 fractionally counted Web of Science indexed core publications in the period 2011–2014 are included in the ranking. | 918 universities are included | 800 universities with at least 200 articles per year published in journals indexed in Scopus, and teaching at least undergraduates in each year during 2010-2014 | In principle all higher education institutions can register for participation. The current version includes about 1,300 institutions. |
| Indicators / dimensions and their weights | • Quality of Education Alumni (10%) Awards (20%) • Quality of Faculty Highly cited researchers (20%) Publ. in Nature, Science (20%) • Research output Publications (20%) • Per Capita Performance (10%) | • Publication counts Articles in English, authored, in core journals • Citation Impact Nr., % Top 1,10, 50 % publications Mean Normalizd Citation Rate • Collaboration Nr, % publ from different institutions Nr, % publ with | • Academic Reputation (40%), based on QS survey • Employer Reputation, based on QS survey (10%) • Faculty Student Ratio (20%) • Citations per Faculty (20%) • International Students (10%) • International Faculty | Performance indicators: • Teaching (30%), mainly based on reputation survey • International Outlook (7.5%) • Research (30%), mainly based on reputation survey • Citations (30%) • Industry Income (2.5%) | Over 30 indicators covering the following main dimensions: • teaching and learning • research • knowledge transfer • international orientation • regional engagement Typical examples of indicators: Quality of teaching (based on survey); citation rate; income from regional sources; nr. spin offs; |

|  |  | geographical collab distance <100 or >5000 km | (10%) |  |  |
|---|---|---|---|---|---|
| Data sources used | Databases on Nobel prizes and field medals;<br><br>Thomson-Reuters Web of Knowledge and Highly Cited researchers; data on academic staff from national agencies | All bibliometric data are extracted from Thomson Reuters' Web of Science | QS Academic Reputation Survey; self-reported data from universities; data from government and other agencies; bibliometric data from Elsevier's Scopus | THE Reputation Surveys; self-reported data from universities; bibliometric data from Elsevier's Scopus | U-Multirank student surveys; self-reported data from universities; bibliometric data from Web of Science and PATSTAT database on patents |